\documentclass[twocolumn]{aastex63}
\usepackage[ruled,vlined,norelsize]{algorithm2e}
\usepackage{amsmath}

\shorttitle{Stellar Variability in the Spectral Domain}
\shortauthors{Zhao, Bedell, Hogg, Luger}

\newcommand{\project}[1]{\textsl{#1}}
\newcommand{\acronym}[1]{{\small{#1}}}
\newcommand{\code}[1]{\texttt{#1}}

\newcommand{\expres}{\acronym{EXPRES}}

\newcommand{\sme}{\acronym{\code{SME}}}

\newcommand{\starry}{\code{starry}}

\newcommand{\phoenix}{\code{PHOENIX}}

\newcommand{\dcpca}{\project{\acronym{DCPCA}}}
\newcommand{\scalpels}{\project{\acronym{SCALPELS}}}
\newcommand{\shell}{\project{\acronym{SHELL}}}
\newcommand{\phiesta}{\project{\acronym{$\Phi$ESTA}}}
\newcommand{\aestra}{\project{\acronym{AESTRA}}}

\newcommand{\cms}{\mbox{cm~s\textsuperscript{-1}}}
\newcommand{\ms}{\mbox{m~s\textsuperscript{-1}}}

\newcommand{\teff}{$T_\mathrm{eff}$}

\newcommand{\logg}{log~$g$}

\newcommand{\dd}{\mathrm{d}}
\newcommand{\obs}{\text{obs}}
\newcommand{\qcol}{orange}
\newcommand{\scol}{blue}
\newcommand{\comrv}{$\beta_{\oplus}$}
\newcommand{\dopvec}{$\vec{I_{\beta}}$}
\newcommand{\intvec}{$\vec{I_{\bullet}}$}
\newcommand{\sftvec}{$\vec{I_{\circlearrowleft}}$}
\newcommand{\dopproj}{$\langle\vec{I_{\beta}},\vec{f}\rangle$}
\newcommand{\intproj}{$\langle\vec{I_{\bullet}},\vec{f}\rangle$}
\newcommand{\sftproj}{$\langle\vec{I_{\circlearrowleft}},\vec{f}\rangle$}
\newcommand{\hamratio}{compact, coherent}

\begin{document}
\title{A Compact, Coherent Representation of Stellar Surface Variation in the Spectral Domain}

\correspondingauthor{Lily L.\ Zhao}
\email{lilylingzhao@uchicago.edu}

\author[0000-0002-3852-3590]{Lily L.\ Zhao}
\thanks{NASA Sagan Fellow} 
\affil{Center for Computational Astrophysics, Flatiron Institute, 162 Fifth Avenue, New York, NY 10010, USA}
\affil{Department of Astronomy \& Astrophysics, University of Chicago, 5640 S Ellis Ave, Chicago, IL 60637, USA}

\author[0000-0001-9907-7742]{Megan E.\ Bedell}
\affil{Center for Computational Astrophysics, Flatiron Institute, 162 Fifth Avenue, New York, NY 10010, USA}

\author[0000-0003-2866-9403]{David W.\ Hogg}
\affil{Center for Computational Astrophysics, Flatiron Institute, 162 Fifth Avenue, New York, NY 10010, USA}
\affil{Center for Cosmology and Particle Physics, Department of Physics, New York University, 726~Broadway, New York, NY 10003, USA}
\affil{Center for Data Science, New York University, 60 Fifth Avenue, New York, NY 10011, USA}
\affil{Max-Planck-Institut f\"ur Astronomie, Königstuhl 17, D-69117 Heidelberg, Germany}

\author[0000-0002-0296-3826]{Rodrigo Luger}
\affil{Center for Computational Astrophysics, Flatiron Institute, 162 Fifth Avenue, New York, NY 10010, USA}

\begin{abstract}\noindent
Time-varying inhomogeneities on stellar surfaces constitute one of the largest sources of radial velocity (RV) error for planet detection and characterization.  
We show that stellar variations, because they manifest on coherent, rotating surfaces, give rise to changes that are complex but useably compact and coherent in the spectral domain.  
Methods for disentangling stellar signals in RV measurements benefit from modeling the full domain of spectral pixels.  
We simulate spectra of spotted stars using \starry\ and construct a simple spectrum projection space that is sensitive to the orientation and size of stellar surface features.  
Regressing measured RVs in this projection space reduces RV scatter by 60-80\% while preserving planet shifts.
We note that stellar surface variability signals do not manifest in spectral changes that are purely orthogonal to a Doppler shift or exclusively asymmetric in line profiles; enforcing orthogonality or focusing exclusively on asymmetric features will not make use of all the information present in the spectra.
We conclude with a discussion of existing and possible implementations on real data based on the presented \hamratio\ framework for stellar signal mitigation.
\end{abstract}


\section{Introduction}\label{sec:introduction}

Stellar surfaces are not spatially or temporally uniform; they differentially rotate, oscillate \citep{kjeldsen1995}, exhibit meridional flows \citep{komm1993}, manifest changing (super-)granulation patterns \citep{dravins1982, rieutord2010}, as well as sport localized features in the form of spots, faculae, and plages \citep{saar1997, meunier2010}.  Observed spectra of stars are a combination of the spectral intensities that arise from these heterogeneous, varying, as well as limb-darkened stellar surfaces \citep{vogt1983, vogt1987}.  Stellar surface changes therefore introduce both shifts to a stellar surface relative to the stellar center of mass (COM) as well as spectral line shape changes that skew measurements of the spectral Doppler shift.

This indicates that there are slightly different possible meanings of ``velocity'' or ``radial velocity'' in precision-measurement contexts.  One possible meaning is the effective Doppler shift of the spectrum, such as what is delivered by optimizing a cross-correlation function (CCF) between an observed spectrum and a mask or template.  This would include all line-of-sight surface velocity variations in the stellar atmosphere.  Another meaning is the latent, not directly observable velocity of the COM of the star relative to the center of mass of the exo-system (i.e., the host star and any orbiting bodies).  This latter velocity, which we denote \comrv, is what is the object of inference in an RV planet search.  Measured RVs will differ from the \comrv\ due to unmodeled stellar signals---a phrase we will use to encompass all stellar surface variations that manifest in the observed spectra---which is currently the largest source of error in extreme-precision RV (EPRV) measurements. The seeming complexity of these signals has necessitated a large body of literature \citep[e.g.][]{jones2017, dumusque2018, gilbertson2020, holzer2020, lafarga2020, rajpaul2020, cretignier2021, collier2021, zhao_jl2022, debeurs2022, lienhard2023} devoted to their characterization and mitigation. An overview and cross-comparison of many of these mitigation methods is given in \citet{zhao2022}.

In this work, we investigate the fundamental properties of stellar signals in the spectral domain, arguing that there should exist a \hamratio\ representation that may be employed to model these signals.  Many methods presented in the literature operate along similar principles by learning such an underlying latent space.  Examples include \dcpca\ \citep{jones2017}, \scalpels\ \citep{collier2021}, and the \shell\ spectral representation \citep{cretignier2022}, which use principal component analysis (\acronym{PCA}) to construct a relevant activity latent space.  Higher dimensional latent spaces may also be represented using neural networks \citep[e.g.][]{debeurs2022, colwell2023, zhao_yn2024} or autoencoders \citep[e.g. \aestra,][]{liang2024}.

Different mitigation methods also vary widely in how center of mass shifts due to planets are preserved.  Methods will first remove planet shifts by centering the data to the best guess \comrv\ \citep[e.g.][]{debeurs2022} or enforce that the modeled variations are orthogonal to the Doppler shift vector \citep[e.g.][]{jones2017}.  Another approach is to construct models that are insensitive to translation shifts, for example by using the autocorrelation function of the data \citep[e.g.][]{collier2019}, moving to the Fourier space \citep[e.g.][]{zhao_jl2020, zhao_yn2023}, or augmenting the data to allow the model to learn what a COM shift looks like \citep[e.g.][]{liang2024}.


In this work we discuss a \hamratio\ framework for representing stellar signals in spectra and explore the efficiency of various possible constraints on models.  Section~\ref{sec:framework} establishes that the observed flux from a star can be written as a linear sum of a finite number of components.  We use this \hamratio\ framework in Section~\ref{sec:data} to simulate spectra arising from a simple spotted star and construct basis vectors sensitive to the amplitude and phase of spots on the stellar surface.  In Section~\ref{sec:constraints} we explore how this framework and other considerations can be used to constrain methods for mitigating stellar signals at the level of the full spectral domain.  We present an example mitigation method building off the \hamratio\ framework in Section~\ref{sec:method}.  We discuss our findings and future directions in Section~\ref{sec:discussion} before summarizing main takeaways in Section~\ref{sec:conclusion}.

\section{Compact, Coherent Variability Framework}\label{sec:framework}

We are motivated by a \emph{multi-component} picture of the stellar surface---i.e., the idea that the surface of a star can be represented by a discretized map of emission intensity that is non-uniform (due to the presence of spots, plages, granulation, and so on).  This idea that the net observed spectrum is a sum over the visible emission map of the stellar surface is closely related to standard Doppler imaging methodologies \citep{vogt1983, vogt1987}, has previously been employed to simulate variable stellar surfaces \citep[e.g.][]{dumusque2014, zhao_yn2023, shahaf2023}, and is the basis of the \starry\ \citep{luger2018, luger2021} framework.  In the \starry\ case, this multi-component picture is expressed as
\begin{equation}\label{eq:surfaceintensity}
I(\lambda_\Omega,\Omega, t) = \sum_k\sum_{\ell m} a_{k\ell m}(t)\,Y_{\ell m}(\Omega)\,I_k(\lambda_\Omega) ~,
\end{equation}
where $I(\lambda_\Omega, \Omega, t)$ is spectral intensity (energy per time per wavelength per solid angle per area), $\lambda_\Omega$ is wavelength in the rest frame of a given stellar surface patch emitting the intensity, $\Omega$ represents the coordinates of the patch on the stellar surface, $t$ is time on the light cone, $a_{k\ell m}$ is the amplitude of spectral component $k$ projected onto spherical harmonic function $Y_{\ell m}(\Omega)$, and $I_k(\lambda_\Omega)$ is the intensity of the spectral component $k$.  
In other words, Equation~\ref{eq:surfaceintensity} is a statement that the intensity emitted at the surface can be modeled as a linear sum of components, where the linear sum depends on position on the stellar surface.

A single observation does not capture the whole stellar surface.  
We consider a tiny visible patch, $\dd\Omega$, of the stellar surface as subtending a tiny solid angular patch, $\dd\omega$, on the sky, with sky coordinates $\omega$ (akin to RA and Dec).
A disk-integrated observation of the star captures the integral of the intensity coming from the visible part of the stellar surface over the sky region $\omega$ covered by the star.
The mapping from the intensity $I(\lambda,\Omega,t)$ at the stellar surface coordinates $\Omega$ to the intensity $I(\lambda,\omega,t)$ at the sky coordinates $\omega$ depends on the orientation of the star.

In addition to this mapping from stellar surface to sky, each solid-angle patch $\dd\omega$ has a different radial-velocity $\beta_\omega(t)$ relative to the star's rest frame because of the (possibly differential) rotation of the star as well as other effects, such as asteroseismic modes.
This leads to a Doppler shift that is a function of sky position:
\begin{align}
    \lambda_\star = \lambda_\Omega\,[1 + \beta_\omega(t)] ~,
\end{align}
that is, the wavelength $\lambda_\star$ in the rest frame of the star is different from the wavelength $\lambda_\Omega$ emitted at the surface, and the shift depends on the position on the surface $\omega$.
Thus the intensity coming from each point $\omega$ on the sky can be written as
\begin{align}
    I(\lambda_\star, \omega, t) &= I(\lambda_\Omega\,[1 + \beta_\omega(t)], \omega, t) \\
    &\approx I(\lambda_\star, \omega, t) + \beta_\omega(t)\,\frac{\dd I}{\dd\ln\lambda}(\lambda_\star, \omega, t)
\end{align}
where $\lambda_\star$ is the wavelength at the stellar rest frame (rather than the surface-patch rest frame wavelength $\lambda_\Omega$), and it is important to remember that by introducing time-varying Doppler shifts of the stellar surface, the mapping from $\Omega$ to $\omega$ now depends implicitly on time as well.
When a star is observed by a spectrograph, what is captured is the flux density $f(\lambda)$, which is an integral of this, or
\begin{align}
    f(\lambda_\obs, t) &= \int I(\lambda_\star, \omega, t)\,\dd\omega \\
    \lambda_\obs &= \lambda_\star\,[1 + \beta_\star(t)] ~,
\end{align}
where there is an additional Doppler shift between $\lambda_\star$ and $\lambda_\obs$ to move from the stellar rest frame to the barycenter rest frame.

The projection from $\Omega$ to $\omega$ is linear, and the Doppler shift can be linearized; therefore, if the original stellar surface intensity $I(\lambda_\Omega,\Omega)$ can be written as a linear sum of components (Eq.~\ref{eq:surfaceintensity}), then the final observed flux $f(\lambda)$ can be written as a linear sum of components
\begin{align}\label{eq:compact}
    f(\lambda_\obs, t) &= \sum_k b_k(t)\,f_k(\lambda_\obs) ~,
\end{align}
where the $b_k(t)$ are coefficients that depend on time (given that the surface is rotating and changing), and the sum is over a larger number of components here than it was above in the surface-intensity assumption.
This story is complicated somewhat by limb darkening and the observed center-to-limb variations in spectral line shape \citep{cavallini1985-02, cavallini1985-09, lohnerbottcher2019, ellwarth2023}. However, there is likely a linearization approach that accurately captures these effects and that keeps the final flux representation low-dimensional.

As an example, we consider the extremely simple (and unrealistic) limit in which there are only two spectral components contributing to $I(\lambda_\Omega, \Omega, t)$---star and starspot, say---and assuming a perfectly linear and geometric projection from $\Omega$ to $\omega$, then the flux $f(\lambda, t)$ can be expressed by a series of components that look something like
\begin{align}\label{eq:components}
    f(\lambda, t) =& \sum_{k=1}^2\sum_{\ell m} a_{k\ell m}(t)\,I_k(\lambda)\,\int \bar{Y}_{\ell m}(\omega)\,\dd\omega \nonumber\\
    &+ \sum_{k=1}^2\sum_{\ell m} a_{k\ell m}(t)\,\frac{\dd I_k}{\dd\ln\lambda}(\lambda)\,\int \beta(\omega)\,\bar{Y}_{\ell m}(\omega)\,\dd\omega ~,
\end{align}
where the $\bar{Y}_{\ell m}(\omega)$ are the spherical harmonics projected onto the sky (projected from $\Omega$ to $\omega$), and we have dropped the ``obs'' subscripts.

The first term in Eq.~\ref{eq:components} for $f(\lambda, t)$ can be thought of as tracing how much of each of the two spectral components project onto the sky (the visible spot fraction, in some sense), and the second term is sensitive to how the derivatives of the two spectral components project onto the Doppler map, $\beta_\omega$, of the rotating stellar surface (i.e., the spot distortion due to the Doppler shift of the stellar surface).
This two-component, purely linear geometric projection model is enormously over-simplified, but it suggests that the spectral data will contain within it detailed information about spot orientation on the stellar surface and the effect of this surface variation on measured Doppler shifts of the spectra.

This derivation suggests that the set of possible spectral variations due to stellar signals is compact---i.e., there's a finite dimensionality to the variations introduced by stellar signals.  In a regression sense, this means that spectral variations due to stellar signals can be represented by a small number of labels.  We note that many existing mitigation methods in the literature implicitly make this assumption by constructing a low-dimensional space to represent stellar variations whether using PCA \citep{jones2017, collier2021, cretignier2022}, a Gaussian-process based regression network \citep{camacho2023}, Fourier basis functions \citep{zhao_jl2020}, or a basis derived from an autoencoder \citep{liang2024}.

Though this is a simple conceptual framework, in practice determining the appropriate/usable representation for stellar signals is a complicated question, as evidenced by the varied approaches in the literature.  The ``best'' representation is likely to depend on the exact geometric projection, the nature of the surface variations in question, and the data properties.  In this work, we focus on the simple, two-component model to demonstrate how this \hamratio\ framework translates to observed spectra and how such a framework can be used to model stellar signals at the spectral level.

\pagebreak
\section{Simulated Data and Projection Space}\label{sec:data}

We simulate observed spectra that are representative of the above framework using \starry, an open-source \code{Python} package that expresses the specific intensity map of a stellar surface as a sum of spherical harmonics \citep{luger2018, luger2021}, as described in Eq.~\ref{eq:surfaceintensity}.  In our simplified model, we incorporate only two spectral components representing a star spot and the otherwise ``quiet'' stellar surface, which we will denote $f_S(\lambda)$ and $f_Q(\lambda)$ respectively.  Using \starry, these two spectral components are assigned to different patches, i.e. $\dd\Omega$, on a coherent, rotating stellar surface.  While the stellar surface does rotate, these toy models do not include either differential rotation or limb darkening.

\subsection{Variations Across the Full Spectral Domain}\label{ssec:spectralDomain}

\begin{figure*}[htb]
\centering
\includegraphics[width=.75\textwidth]{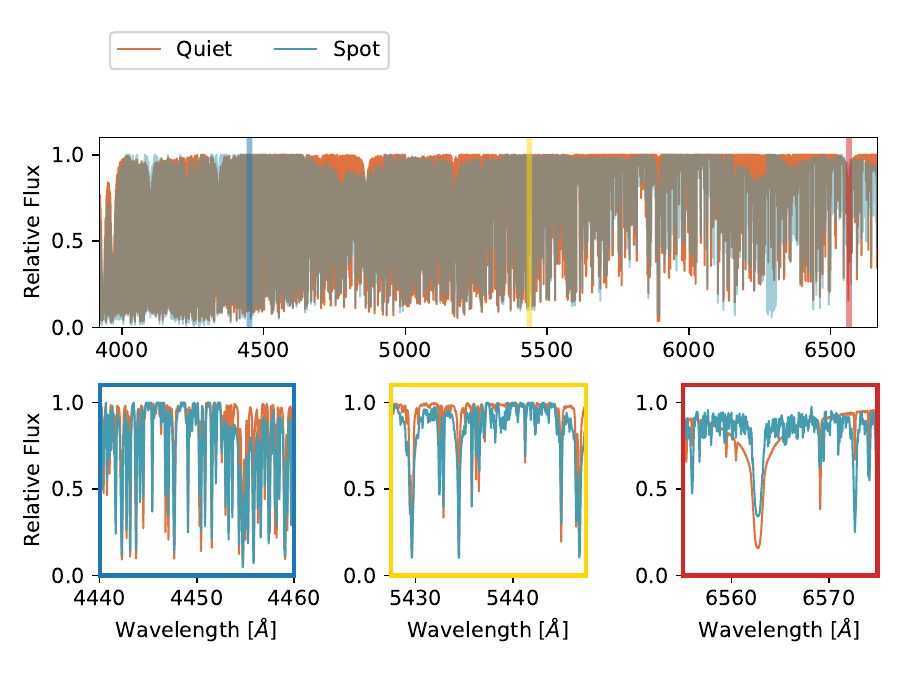}
\caption{Observed spectra of the ``quiet'' (i.e.\ unspotted) solar surface \citep{wallace_quiet_1998} in \qcol\ and observed spectra of the umbra of a sunspot \citep{wallace_spot_2005} in \scol.  \textbf{Top}: the full, optical range of the observed spectrum.  Variations in the quiet and spot spectrum can be seen throughout the wavelength range.  \textbf{Bottom}: Zoom ins on smaller wavelength ranges of the spectra.  Even in regions where it appears the spectra are similar (see the first two subplots, with blue and yellow axes) there are significant deviations.  The right-most plot (red axes) highlights an wavelength region that exhibits large deviations.}
\label{fig:wallace}
\end{figure*}

Resolved observations of the Solar surface confirm that active features, such as spots, give rise to different emergent spectra than the rest of the solar surface, i.e.\ the ``quiet'' surface.  Spectra taken at Kitt Peak of a quiet patch of the solar surface \citep{wallace_quiet_1998} and the umbra of a sunspot \citep{wallace_spot_2005} are overplotted in Figure~\ref{fig:wallace}.  The observed spectrum from the umbra of a sunspot (blue) differs enormously from the quiet solar surface (orange).

Importantly, the spectral differences are not limited to bulk Doppler shifts or line asymmetries---line depth ratios change, new absorption features appear, and the exact nature of this change differs for different spectral lines.  Analysis of the sun spot spectra using Spectroscopy Made Easy \citep[\sme; ][]{valenti1996} found that compared to the quiet spectra, the spot spectra was best fit by a model with significantly lower \teff\ and \logg\ and even a different metallicity (see \cite{komori2023} for more detail).

In the multi-component picture of a star, an observed stellar spectrum integrates over the spectral contributions from across the stellar surface.  The presence of a star spot will therefore contribute a spectrum that is different in complicated ways from the ``quiet'' stellar spectrum in nearly every pixel.  Consequently, the observed integrated spectrum of a spotted star will also be different in every pixel (or wavelength bin) from the integrated spectrum of the same star with no spot.  The nature of these spectral changes will vary on the timescale of the stellar rotation rate, as the spot rotates towards and away from the observer, and the life time of the spot as it evolves.

Given that the entire spectrum changes as a result of stellar surface variations, modeling stellar variability in the full spectrum gives useful and possibly necessary information on the presence and nature of stellar signals.  This is in contrast to commonly used activity indicators (e.g.\ measures of core emission in a single spectral lines) or the cross-correlation function (CCF) of a spectrum with a binary mask that includes only select spectral lines.  These measurements are not optimized to capture the detailed changes due to stellar variability that are present throughout the full observed spectral domain.  Using the full spectra will preserve more information with which to inform models.  However, such a framework also introduces challenges in how to constrain the consequent models across the several thousand pixels and tens to hundreds of observations while ensuring that COM shifts are preserved.  We discuss this further below.

\subsection{Input Data}\label{ssec:modelData}

For our toy models, we assign the quiet and spotted stellar surface high-resolution, synthetic spectra ranging from 5000-7050~nm derived using the \phoenix\ code \citep{hauschildt1999, husser2013}.  The \phoenix\ model of the quiet photosphere, $f_{Q,P}(\lambda)$, makes use of Solar values while the model of a spot, $f_{S,P}(\lambda)$, is tuned to the stellar parameters returned by the best-fit \sme\ model of the umbra of a sunspot \citep{komori2023}.  We note that in using only these two \phoenix\ models, the resultant simulations will not incorporate changes in spectral line shapes that arise from convection, magnetic effects, or center-to-limb variations, so the toy models are limited in these respects.  For our simulations, we degrade the resolution of the \phoenix\ models to R=100,000 with a wavelength spacing of 0.1 \AA\ (corresponding to 300-500 \ms\ in velocity space), which is representative of the resolution of current EPRV spectrographs.

We favored using modeled spectra over observed spectra to mitigate the risk of noise features manifesting as pseudo-spectral-features.  Since only one quiet and one spot spectrum is used to generate all simulated spectra, the noise properties of the input spectra will be inherited by all simulated spectra.  With tens of thousands of pixels in a typical spectrum, it is expected that $5\sigma$ deviations will occur.  Even with high SNR observations, this deviation may manifest at a level relevant to EPRV work.  Such a deviation would then be included in every simulated spectra and could be mistakenly treated as a real feature by analysis methods.  This is especially a danger in the case of data-driven methods.

\subsection{Simulated Systems}\label{ssec:testCases}
\begin{figure*}[ht]
\includegraphics[width=\textwidth]{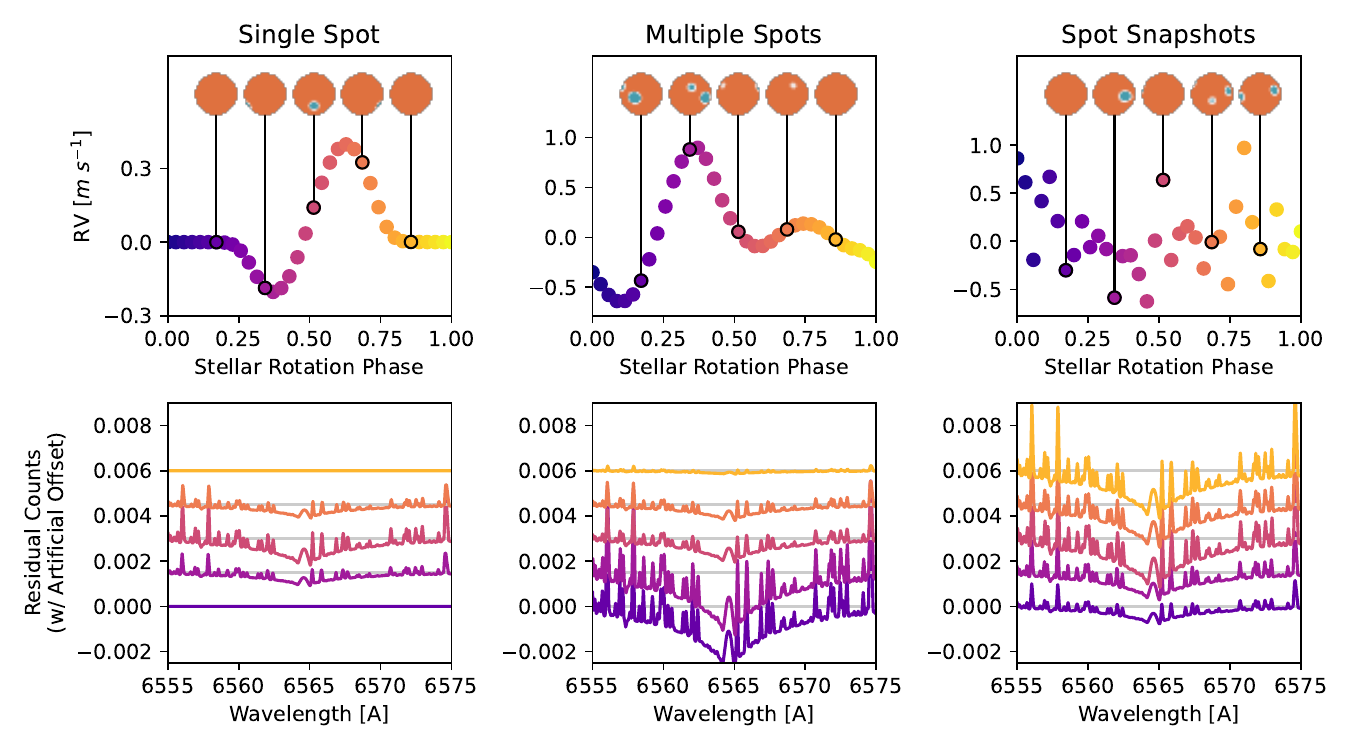}
\caption{Measured RVs (top) and spectral residuals (bottom) for examples of the three different test cases.  Results are shown for the case of a single spot (left), multiple spots (center), and spot snapshots (right), i.e.\ where every spectrum is a completely independent realization of spots.  \textbf{Top}: measured RVs from the simulated spectra.  The top of each subplot shows the observed stellar surface associated with five example phases across the simulated observations.
\textbf{Bottom}: zoomed-in residuals of the simulated spectra relative to the quiet, i.e.\ spotless, spectra for the five example phases in the corresponding color.  The zoom-in is centered on the H-alpha line, which is known to be magnetically sensitive.  Residuals are artificially offset vertically with later phases appearing higher in the plot.}
\label{fig:testCases}
\end{figure*}

We simulated three different spot configurations with the goal of building complexity and testing model dependencies.  The different test cases are:
\begin{enumerate}
    \item \textbf{Single Spot Case}: a time series of spectra across one stellar rotation for a Sun-like star with one spot rotating in and out of view
    \item \textbf{Multiple Spots Case}: a times series of spectra across one stellar rotation for a Sun-like star with multiple spots that rotate in and out of view
    \item \textbf{Spot Snapshots}: a set of spectra for which each spectrum corresponds to a different, random configuration of multiple spots
\end{enumerate}
Measured RVs and spectral changes for an example realization of each of the three cases are shown in Figure~\ref{fig:testCases}.  RVs are measured using a cross-correlation method as implemented in the \expres\ pipeline  \citep{petersburg2020}.  For each test case, we simulate 36 observations spanning a full stellar rotation.

The single spot case represents the simplest realization of RV variations due to a star spot; it is the tutorial level.  We simulate a grid of single spot maps that range in spot latitude and spot radius.  Considering results across the grid allows us to assess how sensitive a formalism is to spot location and size.  The simplicity of this test case allows for more easily interpreted results.

The multiple spot case is a small step towards realism, testing that methods can capture complications that arise with multiple spots.  Five random spots are drawn with latitudes constrained to those typical for sunspots.  Spot sizes are constrained such that the amplitude of the resultant measured RV shift due to the spot presence is on the order of 1-2 \ms.

The spot snapshots case highlights the scenario in which there is no coherence in spot location from one observation to the next.  This test case is therefore sensitive to a method's dependence on time coherence.  The spots for each epoch are selected in the same way as for the multiple spot case.

For each test case, we also include tests with injected Keplerian signals and various levels of white noise.   Two single planet systems are considered.  The first planet has a semi-major amplitude, $K$, of 0.77~\ms\ (on the order of the amplitude of the activity signal) and a period of 0.27 of the stellar rotation period, $P_\star$ (i.e, the planet completes nearly 4 orbits over the course of the simulations, which span one complete stellar rotation).  The second planet has a $K$ of 0.32~\ms\ (on the order of the on-sky precision of current EPRV instruments) with a period of 0.12 $P_\star$.  Simulated spectra are also degraded to SNR of 500, a high but realistic SNR for binned EPRV measurements of bright targets.

\subsection{A Simple Stellar Signal Basis}\label{ssec:vectors}

The power of the proposed framework comes in constructing an appropriate basis vector that captures the relevant spectral changes arising due to stellar surface variability.  In the case of the two-component toy models being considered here, we can use our input spectra, $f_{Q,P}(\lambda)$ and $f_{S,P}(\lambda)$ to directly construct basis vectors sensitive to the visible spot fraction, \intvec, and the spot Doppler distortion, \sftvec.  We use the difference between the quiet and spot spectrum to construct a vector sensitive to the presence of the spot spectrum in data.
\begin{equation}\label{eq:intvec}
    \vec{I_{\bullet}} = f_{Q,P}(\lambda) - f_{S,P}(\lambda)
\end{equation}

We also construct a vector sensitive to spot Doppler distortions by using the difference between the quiet spectrum and the spot spectrum artificially Doppler shifted by some $\beta$.  To ensure the vector is symmetrically sensitive to blue- and red-shifts, we shift the spot spectrum by both $\beta$ and by $-\beta$.  Because we expect these vectors to be approximately opposite to one another, we combine the two vectors by differencing them and dividing by two.
\begin{align}
    I_{\circlearrowleft,+}(\lambda) & = f_{Q,P}(\lambda) - f_{S,P}(\lambda[1+\beta])\\
    I_{\circlearrowleft,-}(\lambda) & = f_{Q,P}(\lambda) - f_{S,P}(\lambda[1-\beta])\\
    \vec{I_{\circlearrowleft}} & = (I_{\circlearrowleft,+}(\lambda)-I_{\circlearrowleft,-}(\lambda))/2
    \label{eq:sftvec}
\end{align}

In the following analysis, we use versions of \intvec\ and \sftvec\ that have been orthogonalized using the Gram-Schmidt algorithm and then normalized.  Enforcing orthogonality in this way ensures that the changes captured by both vectors are preserved while not being replicated.  For our two-component model, this basis should approximately capture the dominant spectral variations due to spots on the surface of the star.

Additionally, we construct the vector $I_{\beta}(\lambda)$ sensitive to pure Doppler shifts \citep[following][]{bouchy2001, jones2017, dumusque2018}.  Projections onto this vector will allow us to assess the relation of various spectral changes to pure Doppler shifts.  Such a vector is necessarily dependent on the spectra being shifted, $f(\lambda)$.  The Doppler shifts of interest, and therefore $\beta$, are small.  We therefore use a Taylor expansion to give an accurate approximation of the shifted spectrum, $f(\lambda[1+\beta])$.
\begin{equation}
    f(\lambda[1+\beta]) \approx f(\lambda)-\beta\lambda \frac{\mathrm{d}f}{\mathrm{d}\lambda}
\end{equation}
The difference between the rest-frame spectrum and a Doppler shifted spectrum is
\begin{equation}
\begin{aligned}
    f(\lambda)-f(\lambda[1+\beta]) &\approx f(\lambda) - [f(\lambda)-\beta\,\lambda\,\frac{\mathrm{d}f}{\mathrm{d}\lambda}]\\
    &=  \beta\,\lambda\,\frac{\mathrm{d}f}{\mathrm{d}\lambda} = \beta\,\left.\frac{\mathrm{d}f}{\mathrm{d}\ln\lambda}\right|_\lambda
\end{aligned}
\end{equation}
From this, we have that the change in intensity expected for a pure Doppler shift will scale linearly with a projection onto \dopvec, defined as
\begin{equation}
\label{eq:dopvec}
    \vec{I_{\beta}} = \lambda \frac{\mathrm{d}f}{\mathrm{d}\lambda}
\end{equation}

In practice, observed spectra are discrete---the intensity is measured at finite wavelength bins---meaning observed spectra are nondifferentiable.  Where relevant, we will take $f$ to represent a smooth approximation of an observed spectrum that has been interpolated to be continuous and from which it is therefore possible to approximate $\frac{\mathrm{d}f}{\mathrm{d}\lambda}$.  The observed, discrete spectra can be treated as a vector, $\vec{f}$.

\begin{figure*}[t]
\centering
\includegraphics[width=.85\textwidth]{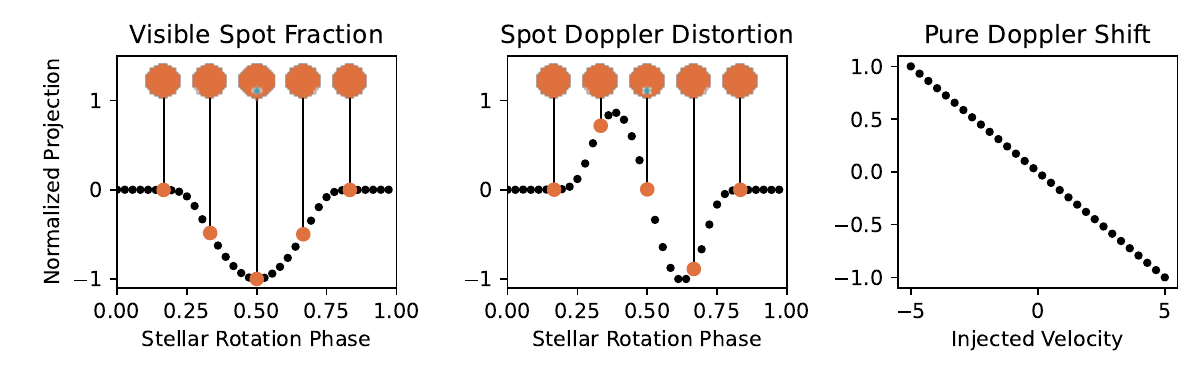}
\caption{Projections onto vectors constructed to be sensitive to visible spot fraction, spot Doppler distortion, and pure Doppler shifts.   \textbf{Left/Center}: normalized projections (y-axis) of simulated spectra of a rotating star with a single spot projected onto either \intvec\ (left) or \sftvec\ (center) as a function of the stellar rotation phase (x-axis).  Images of the visible stellar surface as a function of phase are inset along the top of each subplot.  \textbf{Right}: normalized projections of a spectrum with an injected Doppler shift (x-axis) onto the constructed Doppler vector.}
\label{fig:vectors}
\end{figure*}

We confirm that the constructed vectors are capturing the intended effects by projecting test spectra, $f$, onto each vector.  These projections are expected to scale with the visible spot fraction, \intproj, spot Doppler distortion, \sftproj, and pure Doppler shift, \dopproj.  The resultant projections are shown in Figure~\ref{fig:vectors}.  To test \intvec\ and \sftvec, we project simulated spectra of a single rotating spot into and out of view.  
For the pure Doppler shift vector, \dopvec, we shift a spectrum, $f(\lambda)$ by $\beta = \pm$5 \ms\ and project the residual of the original and shifted spectrum onto \dopvec.

\begin{figure*}[h!t]
\centering
\includegraphics[width=\textwidth]{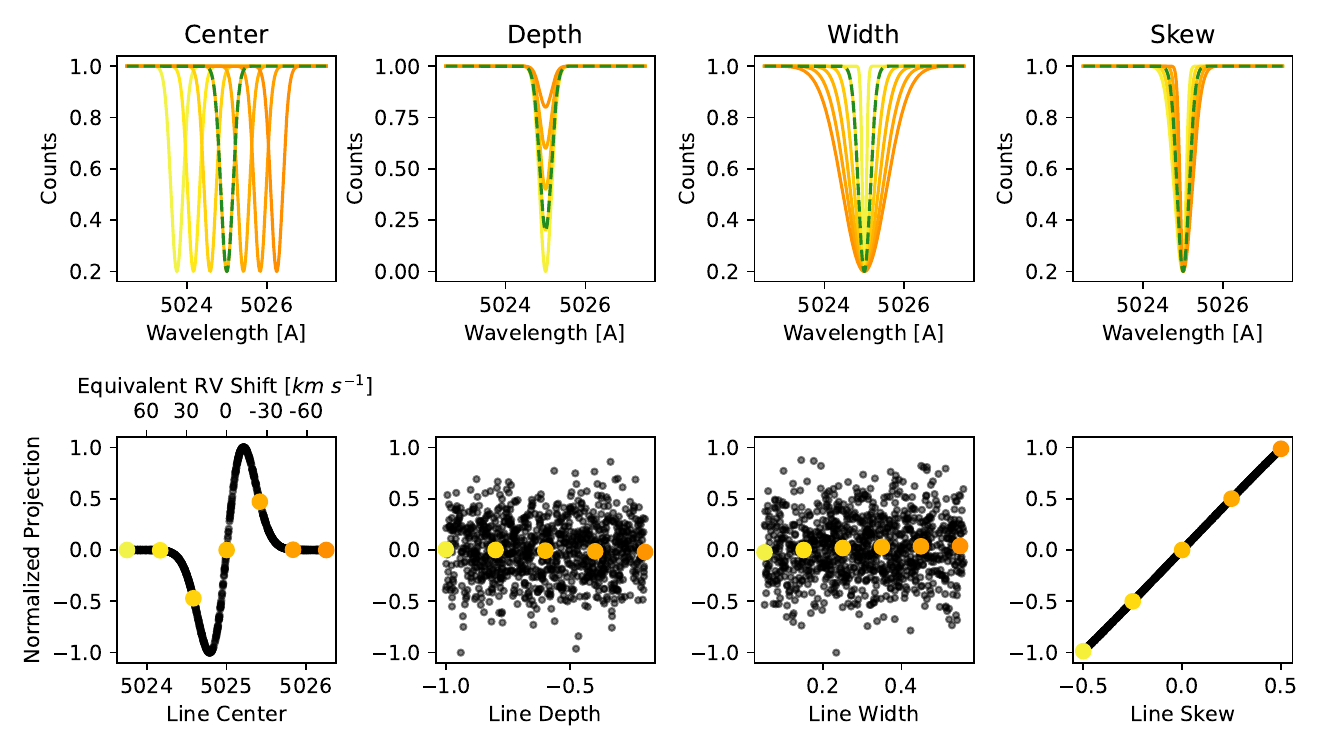}
\caption{\textbf{Top}: changing Gaussian profiles.  From left to right, the Gaussian profiles change in center, depth, width, and symmetry.  To enact a change in symmetry, a bi-Gaussian profile is used.  The dotted, green Gaussian, the same across all four plots, is the reference Gaussian profile used to construct the \dopvec\ against which all other Gaussian profiles are projected.  \textbf{Bottom}: normalized \dopproj\ of different Gaussian profiles with SNR 1000.  The projection of the example profiles shown in the top row with no noise added is shown in the corresponding color in the bottom row.}
\label{fig:dopProj_gauss}
\end{figure*}

\section{Results: Constraints in the Spectral Domain}\label{sec:constraints}

We have simulated full spectra of spotted stars using \starry\ and constructed vectors sensitive to the visible spot fraction, \intvec, the Doppler distortion of spots, \sftvec, and a pure Doppler shift, \dopvec.  We use these simulations and vectors to explore how stellar signals manifest in the spectral domain and what properties can be used to constrain mitigation methods.  Ideally, methods will make use of all available information while preserving the \comrv\ Doppler shift due to orbiting planets.

\hfill
\subsection{Separability}\label{ssec:orthogonality}

One way to ensure that methods preserve the planetary shifts \comrv\ is to constrain the model to capture only variations that are orthogonal to Doppler shift changes \dopvec.  Will this constraint will surely preserve \comrv, it may also render the model insensitive to any spectral variations that are non-orthogonal to a pure Doppler shift.  Using such a constraint introduces an implicit assumption that the effects of stellar signals are either fully separable from a Doppler shift, or, at minimum, that the orthogonal changes are sufficient to capture the effects of stellar signals.

Mitigation methods make use of this category of assumptions in different ways.  For instance, the Doppler-Constrained Principal Component Analysis (DCPCA) method removes the Doppler component from the observed spectra before applying PCA to model the largest remaining variations in the spectra \citep{jones2017}.

Other methods remove translational changes and prescribe all other variation to stellar signals.  \scalpels\ pre-processes input data (CCFs or spectra) with an autocorrelation function, which is invariant to translational shifts, thereby isolating only shape-driven changes in the data \citep{collier2021}.  \phiesta\ decomposes CCFs into Fourier basis functions, where only shape changes will emerge as frequency dependent \citep{zhao_jl2020, zhao_jl2022}.  Such methods will preserve translation shifts, but be insensitive to any stellar signal signatures that mimic a translational shift.

We test the degree to which the expected effects of stellar signals are orthogonal to a Doppler shift by altering the properties of an inverted Gaussian profile (an approximate spectral absorption line) and projecting the different profiles onto the constructed Doppler vector, \dopvec\ (Eq. ~\ref{eq:dopvec}).  The resulting projections are shown in Figure~\ref{fig:dopProj_gauss}.  In each case, \dopvec\ is constructed using a template, non-changing Gaussian profile, shown in the top row of the figure by a dashed green line.  We change the center, depth, and width of the Gaussian profile and introduce asymmetric Gaussian profiles using a bi-Gaussian profile \citep{figueira2013-bgauss}.

Changes in the line center are, as expected, linearly correlated with the Doppler vector until the magnitude of the shift breaks the assumption of a small shift used to construct \dopvec.  Changes in line depth and width are uncorrelated with projection onto the Doppler vector, showing that depth and width changes are orthogonal to a true Doppler shift.  Changes in a line's asymmetry are also correlated with projection onto the Doppler vector.

Modeling variations that are strictly orthogonal to the Doppler vector will therefore capture line depth and width changes, but may not fully capture changing line asymmetries, which are an expected effect of stellar variations.  Removing the translational component of spectral changes will therefore preserve symmetric changes, such as line depth and width, but is likely to affect asymmetric line shape changes.  Though here we are only considering simplistic changes to line shapes, it has previously been shown using time-series observations of $\alpha$ Centauri B that changes due to stellar signals are not strictly orthogonal to the Doppler shift \citep{cretignier2023}.  We further investigate the information content of depth and width changes below.

\subsection{Asymmetry}\label{ssec:asymmetry}

A complementary consideration to enforcing orthogonality to a true Doppler shift (discussed above) ties stellar variations to asymmetric spectral line changes exclusively, as COM shifts are expected to introduce only symmetric line changes.  This assumption underlies the use of skewness in the CCF and line bisectors as activity indicators that gauge the amplitude of line asymmetry present in a spectra \citep[e.g.,][]{queloz2001, dall2006}.

\begin{figure}[hbt]
\centering
\includegraphics[width=.4\textwidth]{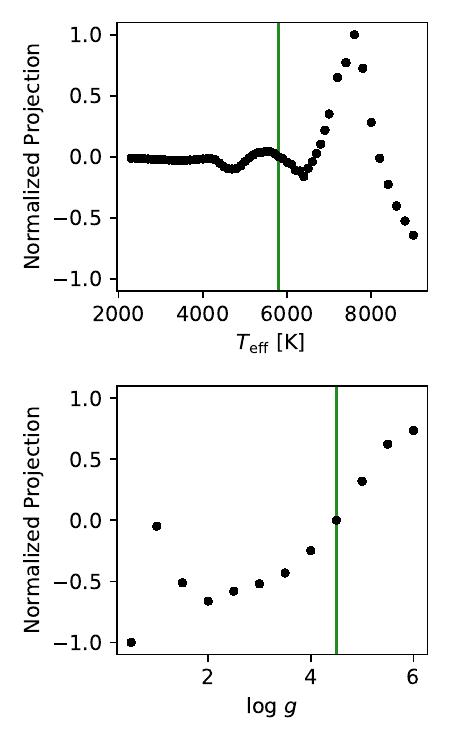}
\caption{Projections onto \dopvec\ of \phoenix\ models with different \teff\ (top) and \logg (bottom) values.  Black points show the projection for simulations for which the spectra were degraded to an SNR of 1000.  Green vertical lines indicate the \teff/\logg\ of the reference simulation used to construct \dopvec.}
\label{fig:dopProj_stellar}
\end{figure}

We test whether stellar variation is exclusively asymmetric by calculating the projection of \phoenix\ models with different stellar properties onto the pure Doppler shift vector.  We have seen that with the Sun---a sunspot emits a spectra with a lower \teff\ and \logg\ among other changes.  Emergent spectra with changes in stellar parameters are therefore expected from stellar surface variation.  Figure~\ref{fig:dopProj_stellar} shows \phoenix\ models at a range of \teff\ and \logg\ values projected onto \dopvec\ constructed using a \phoenix\ model with Solar parameters.  Non-zero \dopproj\ values show that the spectral changes due to changing stellar parameters alone are non-orthogonal to the changes expected from a pure Doppler shift.

Changes in stellar parameters are non-orthogonal to a Doppler shift signal even in the absence of changing line asymmetries.  Because these changes have non-zero \dopproj, they are likely to impact calculated Doppler shifts.  There is therefore a component of stellar surface variations that adds scatter to Doppler shift measurements that would not be captured by measures of line asymmetry alone, requiring the use of additional information (e.g., time correlation, formation properties, etc.).

\subsection{Projected Surface Coherence} \label{ssec:coherence}

We have shown that there should exist a compact representation of stellar spectral variations, and, for our two-component toy model, we have constructed a 2D projection space that traces something akin to the the amplitude of the spot signal (\intproj) and the phase of the spot signal (\sftproj).
A basis that reveals these properties of the spot should enable us to discern the orientation of spots on the surface of a star from the changes in observed spectra.  In other words, the spatial coherence of a spot rotating in and out of view on a stellar surface should trace a coherent surface in the appropriate latent space.  This is expected given that there exists a coherent, rotating stellar surface that underlies all observations.

\begin{figure*}[ht]
\centering
\includegraphics[width=0.35\textwidth]{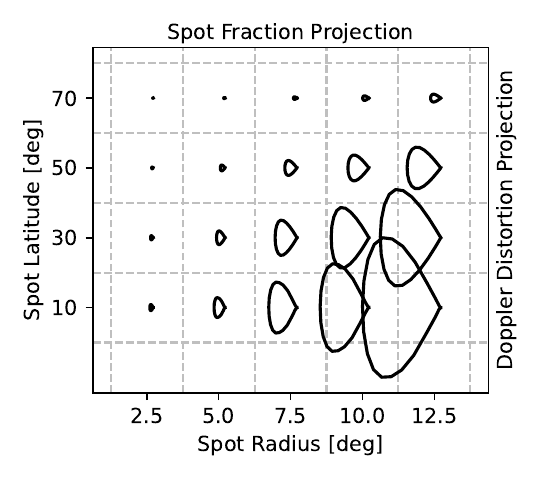}
\includegraphics[width=0.36\textwidth]{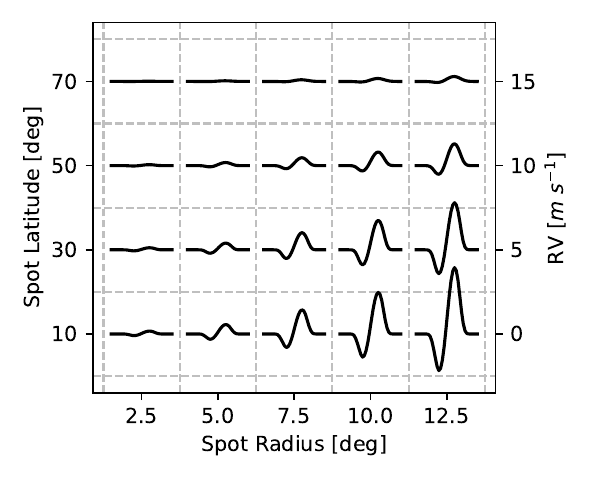}
\caption{Projections (left) and measured RVs (right) of simulated spectra of a rotating star with a single spot.  \textbf{Left}: the \intproj\ (x-axis) and \sftproj\ (y-axis) for each simulation arranged by spot radius (horizontally) and spot latitude (vertically).  Dotted gray lines outline the spot radius/latitude grid.  Each spectrum was projected onto the same vectors so that their relative values are significant.  In this space, a rotating spot forms a coherent loop that varies in shape and size as a function of latitude and spot size.  \textbf{Right}: measured RVs due to the spot where each time series has been artificially shifted to match the same grid layout as the left-hand plot.  The size of the projected loop scales with the amplitude of the measured RV signal.}
\label{fig:heart}
\end{figure*}

We investigate the behavior of a rotating spot in our constructed latent space (i.e., defined by \intvec\ and \sftvec) by projecting simulated spectra of a rotating star with a single spot.  We consider spots with radii ranging from 2.5-12.5$^{\circ}$ and at a latitude of 10-70$^{\circ}$.  Figure~\ref{fig:heart} shows the resultant projection curves for a grid of latitude and radius values in (\intvec, \sftvec) space with artificial offsets in $x$ and $y$.  Here, all spectra were projected onto the same basis vectors.  In other words, though the origins of each curve is offset, the relative amplitudes of the projections can be directly compared.  As predicted, each rotating spot forms a coherent loop.  The size of the loop scales with increasing radius and decreasing latitude, mirroring the amplitude of the resultant measured RV (Figure~\ref{fig:heart} right).

\begin{figure*}[ht!]
\centering
\includegraphics[width=0.675\textwidth]{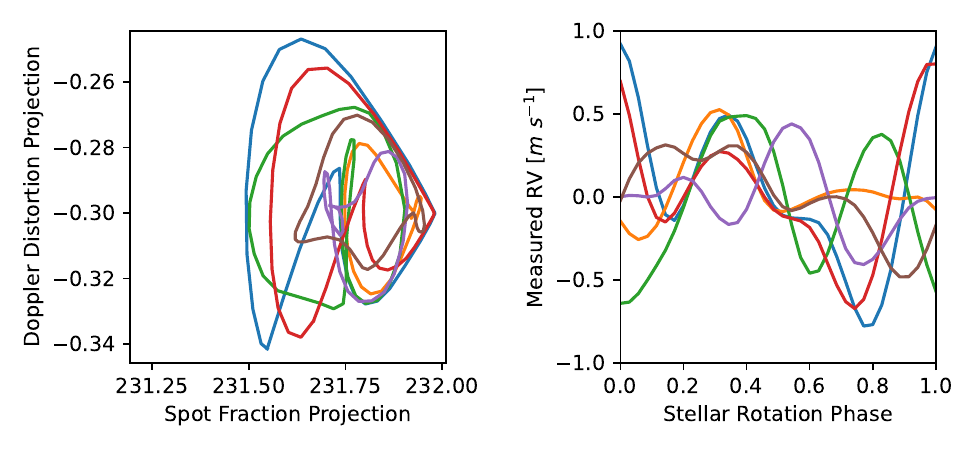}
\caption{Projections (left) onto spot fraction/Doppler distortion vectors (i.e. \intvec\ and \sftvec\ space) and the associated measured RVs (right) for six simulations of a stellar surface with multiple spots.  The color of the projections on the left correspond to the RV time series on the right.}
\label{fig:messyHeart}
\end{figure*}

Projections into the (\intvec, \sftvec) space for simulations of multiple spots are shown in Figure~\ref{fig:messyHeart}.  For six different simulations of stellar surfaces with five randomly generated spots, we show the projections (left) and the measured RV curve (right).  While messier than the clean loops of Figure~\ref{fig:heart} for a single spot, the case of multiple spots still forms a coherent surface.  The deviations can be interpreted as moving between the different coherent loops seen in Figure~\ref{fig:heart} as spots of different size/latitude rotate in and out of view.  The size of the loop continues to scale with the amplitude of the resultant measured RV of the spectra.

\begin{figure*}[t]
\centering
\includegraphics[width=\textwidth]{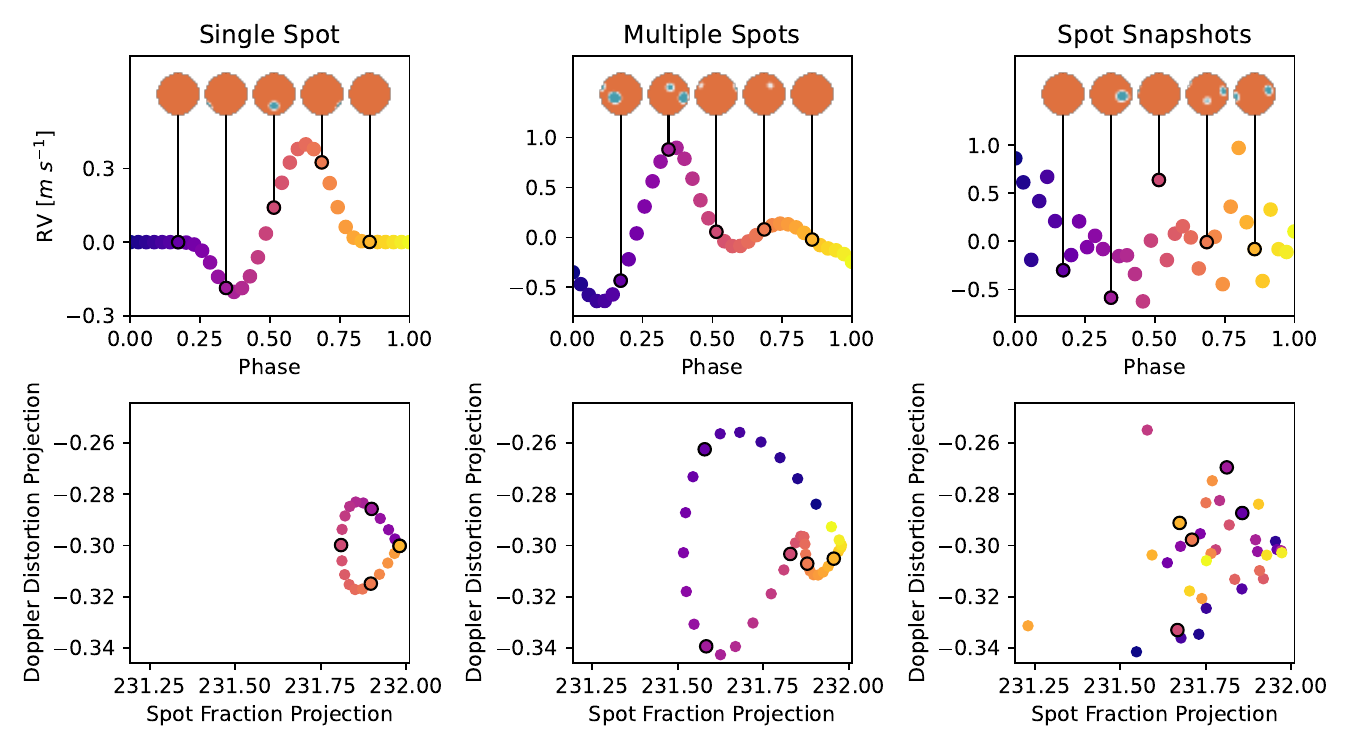}
\caption{Surface maps and measured RVs (top) as well as projections into \intvec\ and \sftvec\ space for the three different test cases.  Points are colored by stellar rotation phase, which allows for the visible stellar surface and measured RV to be mapped to location in projection space.  The projection axes have the same limits across the bottom row of subplots, meaning projections are directly comparable across the bottom row of subplots.}
\label{fig:timeHeart}
\end{figure*}

Figure~\ref{fig:timeHeart} shows measured RV (top) and projections (bottom) for an example of each of the three test cases.  Each observation is shown as a point colored by stellar rotation phase.  The projections across the bottom row of plots uses the same \intvec\ and \sftvec, making the values directly comparable between subplots, where here the focus is the relative value of the projections rather than the absolute value.  For the single spot, the coloring of the points reveal that most of the observations fall at the ``point'' of the projection curve, corresponding to when no spot is in view.  The first circled observation of the multiple spot case (i.e., a dark purple point corresponding to a surface map with no spots) can be seen to fall in a similar area of projection space.

The scatter seen in the projection space for the spot snapshots test case is indicative of each observation tracing an independent surface given that for this test case there is no coherence from one observation to the next.  Despite the scatter, the points all fall in the same area of projection space as would be covered by the full grid of single spot simulations.  The mapping of surface map to projection space and consistency across the test cases confirm that the right basis can capture information about stellar spot orientation from just the integrated light of the star.  Furthermore, this relation does not require temporal coherence.

\subsection{Long-Term Decorrelation of Activity and Planet Signals}\label{ssec:decorrelation}

As an adversarial case, imagine that every instance of an activity state is matched exactly by the orbital position of a planet.  In other words, the phase of the orbiting planet would correspond exactly with an instance of the visible stellar surface.  The spectral variations that arises from the stellar surface changes are therefore exactly matched by the changes due to the planet.  In this case, the effects of the spectral variation and the shift introduced by the orbiting planet will be coincidentally correlated.  Even if we were able to exactly pinpoint a star's rotational phase and variability state (as shown in the previous section), that would not guarantee that we would be able to disentangle the planet signal.

The assumption that the activity and planet signals will be decorrelated is safe for most systems.  Stellar surfaces vary on a range of different time scales and the level of magnetic activity is likely to vary across long-term activity cycles.  Planet periods, on the other hand, are far more stable over longer time spans.  In the long-term, stellar and planet signals will therefore naturally deviate from one another.  The important exception is the case in which the planet is close enough to the star to impact the star's magnetic field, giving rise to stellar variability that is directly correlated with the planet's orbital position \citep{shkolnik2008}.

Particularly with data-driven methods, this property translates to a need for data sets that sample a wide range of combined activity and planet states; they require that multiple planet phases be observed at each variability phase, and multiple variability phases be observed at each planet phase. This is different from a specific requirement on the cadence of observations and is positively correlated with the total amount of data.  Rather than time coherence or volume of data, what is needed is a sampling of disparate activity and planet signals in order to decorrelate the two effects and effectively model the variation from each.

We investigate this constraint further in the below section where we introduce and implement an example mitigation method.

\section{Example Implementation}\label{sec:method}

\begin{figure*}[ht]
\centering
\includegraphics[width=\textwidth]{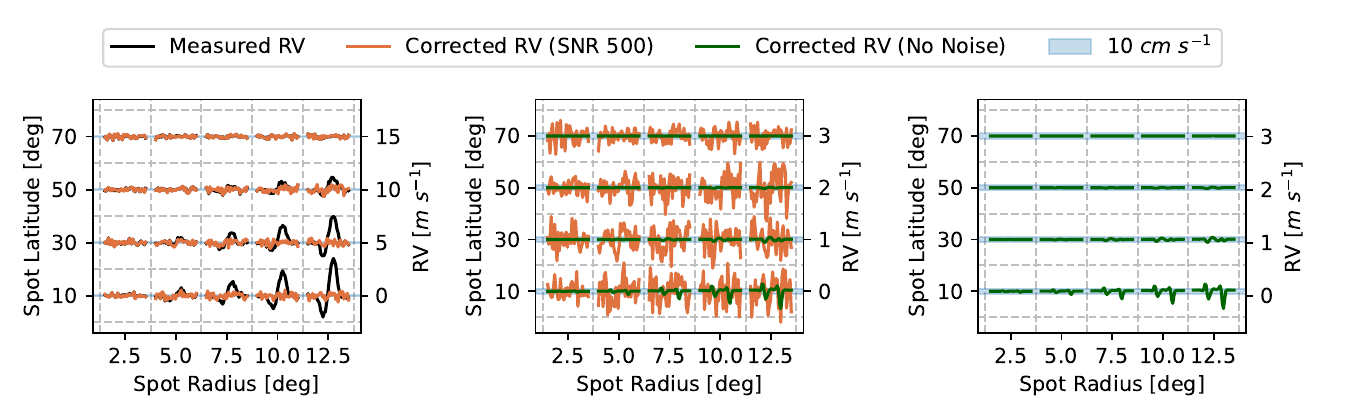}
\caption{RV corrections for simulations with an SNR of 500 of a single spot rotating in and out of view.  Each time series spans exactly one stellar rotation.  For scale, each row is accompanied by blue shading that represents 10 \cms.  \textbf{Left}: measured RVs (black) and corrected RVs (orange) artificially shifted into a grid of spot radius (horizontally) and spot latitude (vertically).  Each row is separated by 5 \ms.  \textbf{Center}: corrected RVs (orange) where each row is now separated by only 1~\ms.  \textbf{Center and Right}: corrected RVs (green) where no noise has been added.}
\label{fig:correctionsSingleSpot}
\end{figure*}

We present the simplest implementation of the \hamratio\ framework described above as a demonstration of how these principles can be used to mitigating stellar signals in spectra.  As discussed in \S\ref{ssec:coherence} and shown in Figure~\ref{fig:heart}, we know that projecting observed spectra onto the constructed spot visibility and Doppler distortion vectors (\intvec\ and \sftvec\ respectively) gives insight onto the spot location and spot size on the stellar surface.  We can therefore use \intproj\ and \sftproj\ to capture and consequently model out the effects of spots on measured RVs.

In this proof-of-concept implementation, we regress the measured RVs against \intproj\ and \sftproj\ to derive an RV correction, $\hat{\beta}$, appropriate for the spot properties revealed by the spectra.  The regression model is fit using a leave-one-out (LOO) method.  The data for which the regression is evaluated on is never used to perform the regression.

The left subplot of Figure~\ref{fig:correctionsSingleSpot} shows the measured RVs (black) and corrected RVs (orange) for a grid of SNR=500 simulations featuring a single spot that ranges in radius and latitude (the same layout as used in Figure~\ref{fig:heart}).  The center subplot of Figure~\ref{fig:correctionsSingleSpot} shows a zoom-in on the corrected RVs (orange) with blue shading that spans $\pm$0.1~\ms, the approximate RV amplitude of an Earth-like planet orbiting a Sun-like star, to help provide scale.  The measured RVs and orange corrected RVs are shown for data with a simulated SNR of 500.

The center and the right-most subplot of Figure~\ref{fig:correctionsSingleSpot} shows corrected RVs in green for the case where no noise has been added.  With the scatter from noise removed, it is revealed that the amplitude of the corrected RVs scales with the amplitude of the measured RVs.  The corrected RVs also appear to trace the derivative of the measured RVs.

\begin{figure*}[h]
\centering
\includegraphics[width=\textwidth]{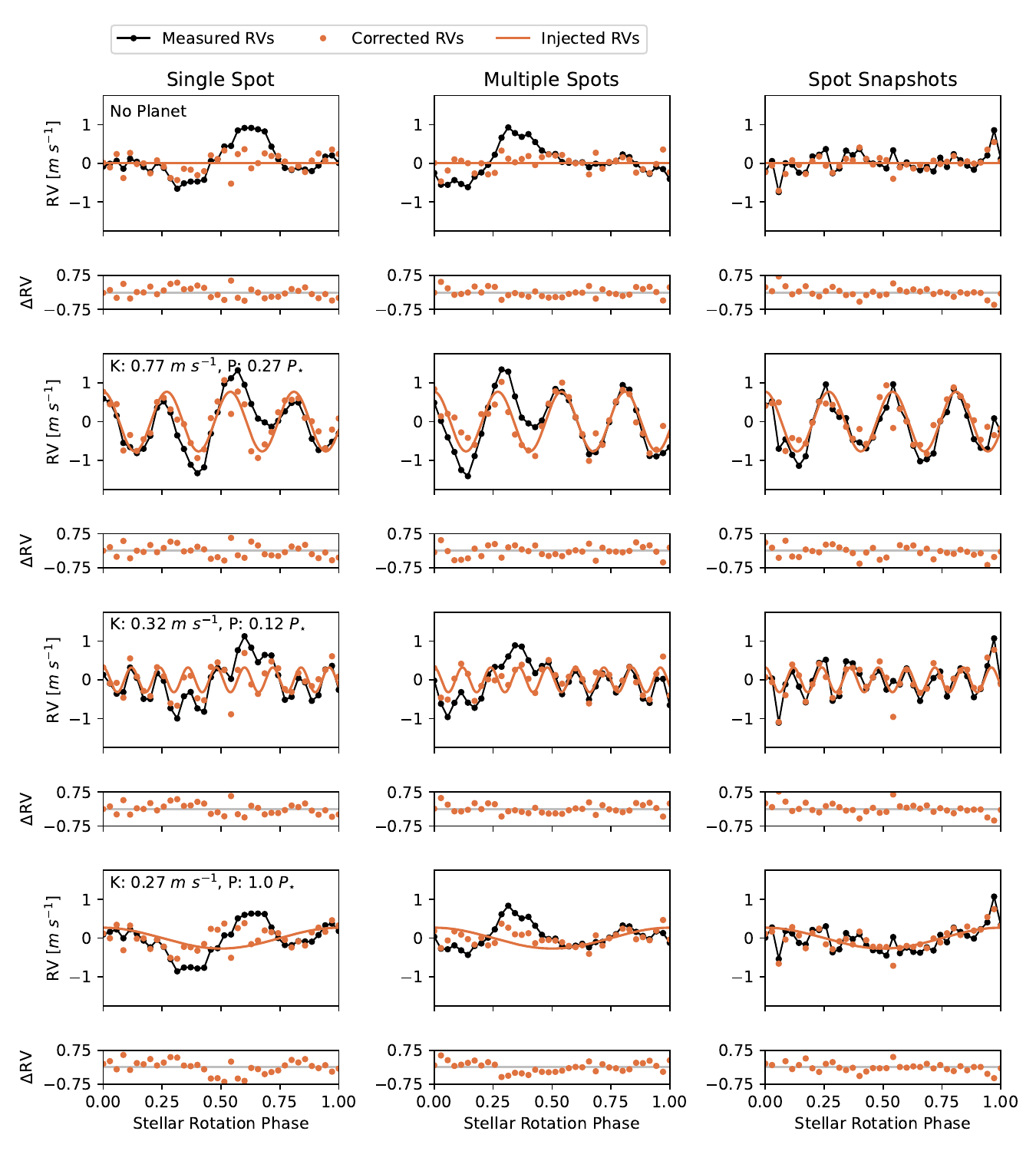}
\caption{Corrected RVs for a random example of each of the three different test cases (corresponding to each column) for different injected planetary systems (Corresponding to different rows) at SNR 500.  Each plot shows the measured RVs (black), the corrected RVs (orange points), and the injected RV curve (orange solid line).  The residuals between the corrected and injected RVs ($\Delta RV$) are shown below each plot.  The first row shows the case of no planet; in the next three rows, the semi-major amplitude $K$, and period $P$, of the injected planetary signal is given in the top-left corner of the left-most plot in the row.}
\label{fig:corrections}
\end{figure*}

Figure~\ref{fig:corrections} shows the measured (black points and line) and corrected (orange points) RVs for an example of each of the three test cases: a single spot, multiple spots, and random spots (as described in \S\ref{ssec:testCases}).  Simulations are degraded to an SNR of 500.  The orange curve in each plot shows the injected pure Doppler shifts with the residuals between this orange curve and the corrected RVs shown below.  If the method is performing as desired, the corrected RVs will fall on the injected RV curve.

Corrected RVs fall close to zero in the case of no planet and trace the planet curve when a Keplerian signal is injected, even for a planet amplitude of 0.32~\ms\ (third row).  The last row shows the case in which the planet period and stellar rotation period are the same.  In other words, the planet signal and stellar rotation signal are completely correlated.  For the case of the correlated planet and stellar signal (last row), we see that the corrections perform more poorly in the single spot and multiple spot case.  As discussed in Section \ref{ssec:decorrelation}, this is expected from the fundamental information content of the data.  We note that this is not true in the case of spot snapshots (right-most column), for which the phase of the star has no relation to the stellar surface, effectively breaking the correlation.

\subsection{Comparison to Common Activity Indicators}\label{sec:indicators}

\begin{figure*}[ht]
\centering
\includegraphics[width=\textwidth]{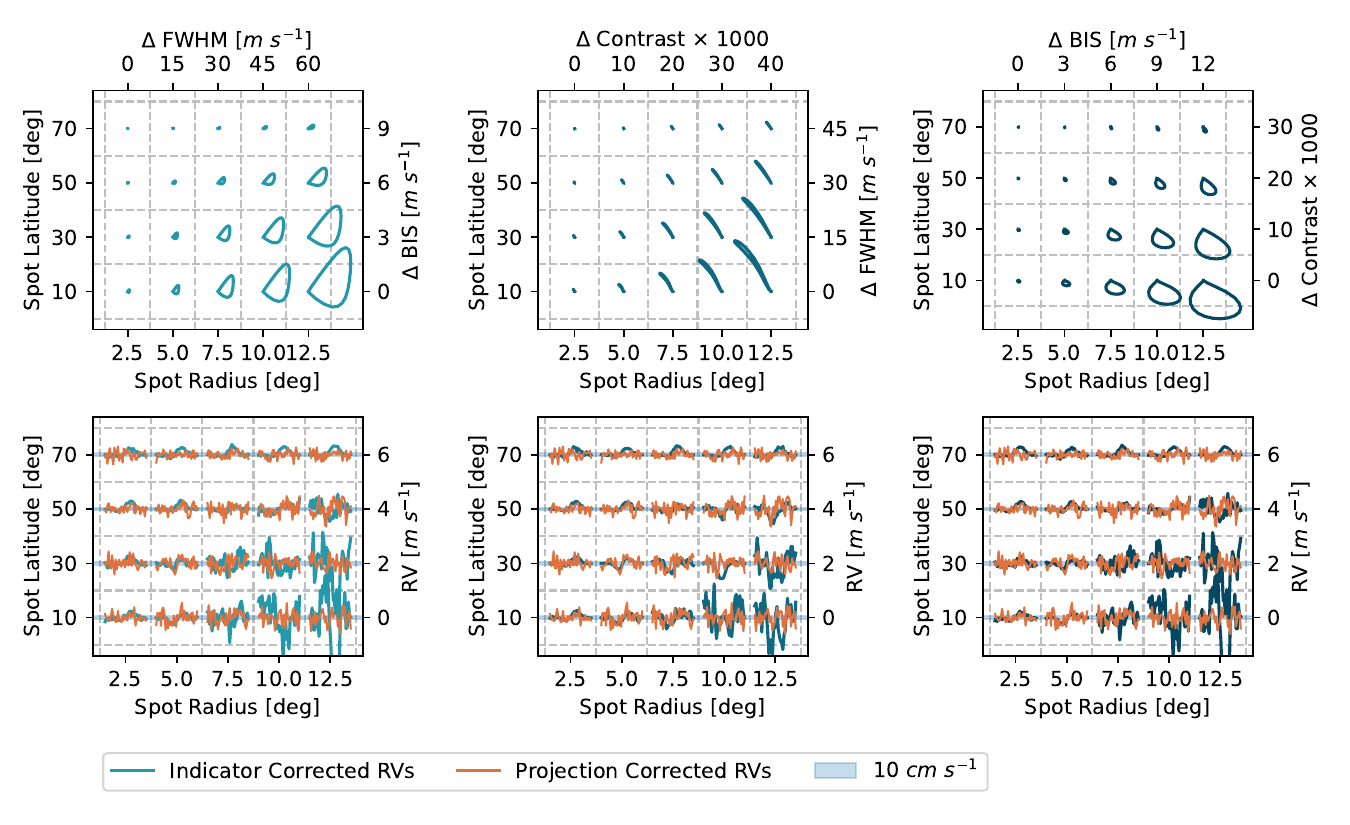}
\caption{\textbf{Top:} Common activity indicators measured from simulated spectra of a rotating star with a single spot over a grid of spot radii and latitudes (similar to Figure~\ref{fig:heart} left).  The two indicators being compared and their relative values are given to the top and right of each subplot.  The rotating spot spectra form a similar loop in indicator space as in the simplistically constructed \intvec\ and \sftvec\ projection space used in this work.  \textbf{Bottom:} RVs detrended relative to the measured indicators for each pair of indicators in shades of blue (similar to Figure~\ref{fig:correctionsSingleSpot}).  For comparison, RVs corrected with respect to projections onto \intvec\ and \sftvec\ are shown in orange.}
\label{fig:indicators}
\end{figure*}

This simplistic linear-regression implementation of the constructed \hamratio\ framework approximates how activity indicators were historically employed in the field.  Indicators such as the FWHM of the CCF, the CCF contrast (i.e.\ depth), and the CCF Bisector Inverse Slope \citep[BIS,][]{queloz2001} also provide measures of the amplitude of stellar signals present in a RV measurement.  Stellar signals were mitigated by detrending measured RVs against activity indicators.

We investigate how these activity indicators relate to the projection space constructed here by using CCF FWHM, contrast, and BIS in place of projections onto \intvec\ and \sftvec.  Figure~\ref{fig:indicators} plots pairs of indicators calculated from simulated spectra of a grid of single rotating spots (mirroring Figure~\ref{fig:heart} left) and the consequent corrected RVs (similar to Figure~\ref{fig:correctionsSingleSpot}).  These corrected RVs can be compared to the RVs corrected using \intproj\ and \sftproj, which are overplotted in orange.

We see that plotting indicators against one another also form coherent, closed loops.  This is expected as the constructed projections and the activity indicators are both intended to capture the amplitude of spot features.  FWHM and contrast more closely follow \intproj\ while BIS is more similar to \sftproj.  The loops in indicator space are somewhat more skewed, most noticeably in FWHM vs. contrast space, as the effects captured by these indicators are not strictly orthogonal to one another.

\subsection{RMS of Corrected RVs}

\begin{figure*}[ht]
\centering
\includegraphics[width=.9\textwidth]{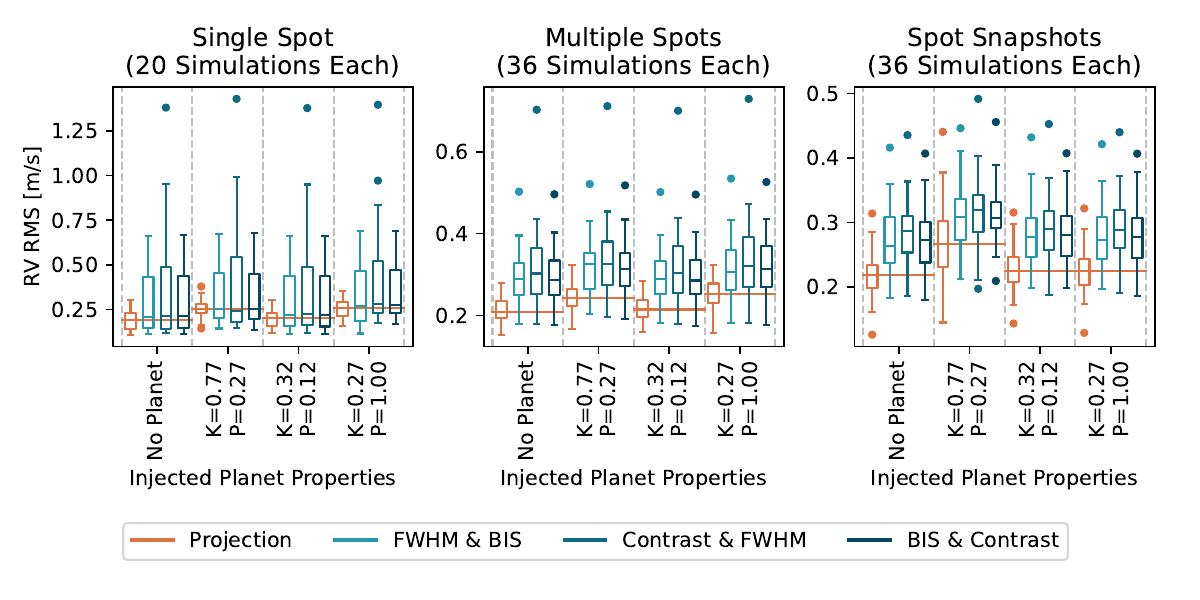}
\caption{Distributions of the RV RMS after correcting against the constructed projection space (i.e.\ \intvec\ and \sftvec\ space) in orange and correcting against activity indicator pairs in blue.  Each box-and-whisker plot represents the RV RMS distribution across all simulations, where the total number of simulations is given above each sub-plot.  Different box-and-whisker plots represent different injected Keplerian signals; the properties of the injected planets are given along the x-axis with semi-major axis $K$ given in units of \ms\ and period $P$ given in units of stellar rotation.  The median RV RMS for corrections against \intvec\ and \sftvec\ is drawn across the set of simulations corresponding to each injected planet.}
\label{fig:correctionsDistribution}
\end{figure*}

Figure~\ref{fig:correctionsDistribution} shows the distribution of the corrected RV RMS for the three different test cases (corresponding to the three subplots) and different injected planetary signals.  Distributions are shown for RVs that were corrected using the constructed \intvec\ and \sftvec\ space (orange) as well as RVs corrected using different pairs of activity indicators (shades of blue).  Simulations are degraded to an SNR of 500.  The distributions encapsulate (left) a grid of 20 single spots\footnote{Four different spot latitudes; five different spot radii.}, (center) 36 different multiple spot maps, and (right) 36 different realizations of spot snapshots.  Distributions are shown for the same four planetary systems (including no planet) as shown in Figure~\ref{fig:corrections}.

We see that the RVs corrected using \intproj\ and \sftproj\ are consistently reaching a median RMS of 20 to 30 \cms.  While the RMS of RVs detrended against pairs of indicators is similar in the case of a single spot (left-most plot), the RMS is higher in the more realistic case of multiple spots.

Even with a very simple implementation, this correction is successfully reducing stellar signals to around 20~\cms\ levels as well as preserving planet signals.  Here we are using only two components.  By intentionally constructing components to be sensitive to imprints of stellar signals across the spectral range, we recover lower RV RMS than using classic activity indicators.  It is likely that incorporating more components will result in better corrections.  Considering that the corrected RVs for the single-spot, no-noise case (Figure~\ref{fig:correctionsSingleSpot} right) trace the derivative of the measured RVs, it is possible that a higher-order model is needed to fully capture the stellar signal.

\section{Discussion}\label{sec:discussion}

We have demonstrated that spectral changes due to stellar surface variations are expected to have a compact representation in the spectral domain. This compactness can be used to model out stellar signals while preserving Doppler shifts from orbiting planets.  Building on the multi-component picture of a stellar surface as a map of emission intensity (Eq.~\ref{eq:surfaceintensity}), we showed that the final observed flux of a star can be linearized and represented by a finite sum of components (Eq.~\ref{eq:compact}).  We used the multi-component stellar surface picture to simulate test spectra of spotted stars with \starry.  We then constructed a set of vectors sensitive to visible spot fraction (\intvec, Eq.~\ref{eq:intvec}) and the Doppler distortion of a Spot (\sftvec, Eq.~\ref{eq:sftvec}).

These vectors define a projection space that captures coherent surfaces corresponding to spot orientation on a rotating star using just spectral changes (Figures~\ref{fig:heart} - \ref{fig:timeHeart}).  Regressing measured RVs in this projection space corrected for spurious RVs due to a spotted stellar surface while preserving planet signals (Figures~\ref{fig:corrections}).  While existing activity indicators approximated this projection space (Figure~\ref{fig:indicators}), by intentionally constructing an activity-sensitive projection space that makes use of the entire spectrum, we were able to achieve a greater reduction in RV RMS (Figure~\ref{fig:correctionsDistribution}).  More advanced ways of constructing an N-dimensional latent projection space will likely capture more of the spot properties that are imprinted in observed spectra.  The conceptual framework presented here can even be seen as something like a baby step towards spectroscopic imaging of a full stellar surface.

In general, methods attempting to disentangle stellar variation from planetary shifts have much to gain from working at the full-spectrum level.  We showed in \S\ref{ssec:spectralDomain} that stellar surface variations give rise to changes across the spectral domain, providing a wealth of information.  We investigated various possible constraints at the spectral level.

We found that enforcing orthogonality to a Doppler shift or modeling only asymmetric features will not capture all spectral changes due to stellar variability.  Such constraints may play a necessary role in the case of very little data, but are likely unnecessary with large amounts of data.  With more data, there is more information available on how spectra change in response to stellar variability, negating the need to enforce artificial constraints on the spectral changes being modeled.

If the effect of stellar surface variability shares many similarities with a true Doppler shift, it will likely be necessary to use a combination of spectral and time-domain information to separate stellar and planetary signals.  For example, time correlation between a stellar signal and measured RVs could be what is needed to separate out the component of stellar signals that is parallel to stellar surface variability.  Implementing such a method will put time requirements on the collected spectra.

Building information for data-driven methods is dependent on collecting data that captures a range of stellar variability and planet states in order to successfully decorrelate the two.  The range in states is what is needed to construct a compact model that captures both the range of possible stellar surface variations and preserves \comrv\ from planets.  Because stellar variations exist on a stellar surface, there is an expected coherence in this projection space that could be leveraged to inform mitigation methods.  With the appropriate projection space, it will even be possible to extrapolate properties of the stellar surface from the integrated spectra alone.

\subsection{Recommendations for Method Testing}

We use \starry\ to simulate full spectra of a variable stellar surface.  For the toy models used here, we considered only two components: the quiet stellar surface and a star spot spectrum.  A natural next step is moving to more complicated stellar surfaces that incorporate, for instance, different components for plage and faculae clusters, alternative spot components, and activity features that evolve.  Different oscillation states can be introduced via the surface Doppler map $\beta_\omega(t)$.  Changing granulation patterns could be captured with different ``patches" ($\dd\Omega$) representing different granules, though the resolution of a typical granulation presents some issues for the spherical-harmonic based \starry\ code.

Our simulated test spectra were constructed using \phoenix\ spectra with different stellar parameters as opposed to real data.  Using real data carries the risk of misinterpreting noise, even white noise, as a true astrophysical signal.  This concern must be taken into account at the level of EPRV work.  On the other hand, simulated data lack needed realism; \phoenix\ models do not reflect line shape changes due to shifting convection patterns and the simplistic models used here neglected limb-darkening effects as well as center-to-limb line variations.  Such effects are likely to impact a model's ability to disentangle stellar signals from pure Doppler shifts. Using real observations could, for example, result in better informed models by providing additional spectral imprints of stellar variability not captured by simulations.  We have not explored such changes in our simple model as we choose to keep the model simple and thereby interpertable.  To fully inform method development and performance, there is a need for both more realistic simulations of full spectra as well as more high-fidelity observed spectra capturing stellar variability across a range of spectral types.

To test the \hamratio\ framework, we simulated stellar surfaces with (1) a single spot, (2) multiple spots, and (3) random realizations of spots, or ``spot snapshots."  These different test cases provided (1) interpretability, (2) increased complexity, and (3) independence from time coherence respectively.  We encourage other proposed mitigation methods to run on a similar set of simulated stellar surfaces.

We injected Keplerian shifts to the simulated spectra at the wavelength level to ensure such shifts can be preserved.  We found that it was important to test a range of planet-to-stellar signal amplitude ratios.  Planet periods that are the same as or integer multiples of the stellar rotation should be avoided as the planet and stellar signals will be more correlated in the resultant data.  Fitting models on a different subset than the data the model is evaluated on should be standard practice to protect against overfitting.

\subsection{Future Implementations}\label{ssec:fullImplementation}

Moving to a full implementation suitable for real observations will require careful thought on how to best learn the underlying projection space.  This construction could be motivated by physics and/or invoke data-driven modeling.  Here we demonstrated the potential power of such a projection space, but constructed this space using known spectra of the quiet and active regions of the simulated spotted star.  While we may never know the true intrinsic quiet/active spectra for real stars, it is possible that approximations using spectral models with different stellar parameters may be sufficient for defining a usable projection space.  It is also worth exploring whether this projection space can be derived using the data itself, such as with clever constraints on PCA-related techniques.  Such considerations are out of scope for this initial paper, but will be the focus of future work.

The definition of the projection space could additionally incorporate spectral information beyond intensity as a function of wavelength.  For example, information on the formation temperature of absorption features could be included \citep{almoulla2022}.  Imposing time coherence for suitable data sets could also augment the information gained from the spatial coherence of the stellar surface.
    
In order for the proposed framework to remain applicable to real data, the dominant changes in the observed spectra must be due to stellar signals.   The presence of significant spectral variation that is non-orthogonal to the changes expected from stellar variation may bias the resultant model.  This requires that instrument systematics be negligible relative to the expected changes due to stellar variation.  As one example, we found continuum normalization to have a large effect on results of the example implementation.  Other data properties, such as requisite SNR, resolution, etc.\ are likely to depend specifically on the implementation in question.

The most useful implementation of the discussed constraints will be inherently tied to the specifics of the data and system at hand.  The expected dimensionality of the latent projection space is likely to change with the relative amplitude of the stellar and planet signals.  The information content of the available data---which changes with properties such as SNR, wavelength range, sampling of planet/star phases, and so forth---may limit the possible options for constructing the latent space.  For instance, PCA will be easier to implement on higher SNR data and/or systems with a higher activity to planet signal ratio.  The number of relevant components may also scale with this activity/planet signal ratio or with the activity level of the star.

Stellar signals must be mitigated to sub-meter-per-second levels in order to achieve EPRV measurements in an astrophysically useful sense.  Pursuing this issue at the spectral level grants access to more information with which to capture and understand stellar surface variations.  This paper explored different constraints that such models can use in terms of the amount of information preserved and the flexibility allowed.  We showed that spectral variations from stellar signals lie in a compact space and used this property to inform an example mitigation method.  The proposed framework is not yet a full method that works in general.  This contribution is a conceptual one that can form the basis of a new class of methods for both next-generation and legacy data sets.

\section{Conclusions}\label{sec:conclusion}

We have shown through derivation that our basic assumption of a multi-component stellar surface means there must exist a compact representation of how stellar surface variations manifest in spectra (Eq.~\ref{eq:compact}). This \hamratio\ representation can be used to model out the effects of stellar signals while preserving planet shifts (\S\ref{sec:method}). An appropriate choice of projection space may even enable spectroscopic Doppler imaging, revealing the surface features of a star from spectra (\S\ref{ssec:vectors}, \ref{ssec:coherence}).

In investigating how stellar surface variations manifest in the spectral domain, we found that:
\begin{itemize}
    \item The full spectral format should be used to preserve the greatest amount of information on stellar surface variations (\S\ref{ssec:spectralDomain}).
    \item Considering effects that are only asymmetric (\S\ref{ssec:asymmetry}) or orthogonal to a pure Doppler shift (\S\ref{ssec:orthogonality}) may be excluding available information on stellar signals.
    \item Though related to high cadence or large quantity of data, the information content of data sets builds most directly with the number of different activity/planet states captured (\S\ref{ssec:decorrelation}).
    \item The ``best'' mitigation method is likely to depend on data quality, quantity, and system specifics (e.g., relative amplitude of the planet and activity signals).
\end{itemize}

We found the following useful principles for simulating data and testing mitigation methods:
\begin{itemize}
    \item \code{Starry} can be used to simulate full spectra of tunable stellar surface variations (\S\ref{sec:data}).
    \item Caution is needed when using small numbers of real observations (as opposed to theoretically modeled data) as input to simulations (\S\ref{ssec:modelData}).
    \item Testing methods on spot snapshots (where the spot configuration is completely different from one observation to the next) helpfully probes a method's sensitivity to many realizations of multiple spots as well as its dependence on time coherence.
    \item Method tests should include the ability to preserve injected Keplerian signals with a range of planet-to-activity amplitude ratios and planet-to-activity period ratios.
    \item Methods should never be trained on the same data for which they will be evaluated to protect against overfitting.
\end{itemize}

While we do not implement a fully-fledged mitigation method on real data in this work, we offer some guidelines and thoughts on doing so in the future (\S\ref{ssec:fullImplementation}). The basic properties of how stellar variability manifests in spectra that we have demonstrated in this paper may also be used to understand the strengths, weaknesses, and underlying assumptions of other activity mitigation methods in the literature. Stellar variability at the spectral level is a formidable problem for EPRV science, but we have shown in this work that these stellar signals may be parameterized as a \hamratio\ representation in the data, offering a path forward for the future.

\software{SciPy library \citep{scipy}, NumPy \citep{numpy, numpy2}, Astropy \citep{astropy2013,astropy2018}, \starry\ \citep{luger2018,luger2021}.}

\hfill
\vspace{-30pt}
\acknowledgements
It is a pleasure to thank
Andy Casey (Monash),
John Brewer (SFSU),
Claire Komori (SFSU), 
and the \textsl{Terra Hunting} Science Team for valuable discussions.  
The Flatiron Institute is a division of the Simons Foundation.  Support for this work was provided by NASA through the NASA Hubble Fellowship grant HST-HF2-51569 awarded by the Space Telescope Science Institute, which is operated by the Association of Universities for Research in Astronomy, Incorporated, under NASA contract NAS5-26555.

\bibliography{paper}

\end{document}